\documentclass[%
 reprint,
superscriptaddress,
showpacs,
 amsmath,amssymb,
 aps,
prx,
]{revtex4-1}

\usepackage{graphicx}
\usepackage{dcolumn}
\usepackage{bm}
\usepackage{tabstackengine}
\stackMath

\newcommand{\bra}[1]{\ensuremath{\left\langle#1\right|}}
\newcommand{\ket}[1]{\ensuremath{\left|#1\right\rangle}}
\newcommand{\bracket}[2]{\ensuremath{\left\langle#1 \vphantom{#2}\middle|  #2 \vphantom{#1}\right\rangle}}

\newcommand{\yb}[0]{\ensuremath{^{171}\text{Yb}^+}}

\begin{document}

\preprint{APS/123-QED}

\title{Generation of high-fidelity quantum control methods for multi-level systems}

\author{J. Randall}
\affiliation{Department of Physics and Astronomy, University of Sussex, Brighton, BN1 9QH, UK}
\affiliation{QOLS, Blackett Laboratory, Imperial College London, London, SW7 2BW, UK}
\author{A. M. Lawrence}
\affiliation{Department of Physics and Astronomy, University of Sussex, Brighton, BN1 9QH, UK}
\affiliation{QOLS, Blackett Laboratory, Imperial College London, London, SW7 2BW, UK}
\author{S. C. Webster}
\affiliation{Department of Physics and Astronomy, University of Sussex, Brighton, BN1 9QH, UK}
\author{S. Weidt}
\affiliation{Department of Physics and Astronomy, University of Sussex, Brighton, BN1 9QH, UK}
\author{N. V. Vitanov}
\affiliation{Department of Physics, St. Kliment Ohridski University of Sofia, 5 James Bourchier blvd, 1164 Sofia, Bulgaria}
\author{W. K. Hensinger}
\thanks{Author to whom correspondence should be addressed. w.k.hensinger@sussex.ac.uk}
\affiliation{Department of Physics and Astronomy, University of Sussex, Brighton, BN1 9QH, UK}

\date{\today}

\begin{abstract}
In recent decades there has been a rapid development of methods to experimentally control individual quantum systems. A broad range of quantum control methods has been developed for two-level systems, however the complexity of multi-level quantum systems make the development of analogous control methods extremely challenging. Here, we exploit the equivalence between multi-level systems with SU(2) symmetry and spin-1/2 systems to develop a technique for generating new robust, high-fidelity, multi-level control methods. As a demonstration of this technique, we develop new adiabatic and composite multi-level quantum control methods and experimentally realise these methods using an $\yb$ ion system. We measure the average infidelity of the process in both cases to be around $10^{-4}$, demonstrating that this technique can be used to develop high-fidelity multi-level quantum control methods and can, for example, be applied to a wide range of quantum computing protocols including implementations below the fault-tolerant threshold in trapped ions.
\end{abstract}

\maketitle

\begin{figure*}
\centering
\includegraphics[width=1.7\columnwidth]{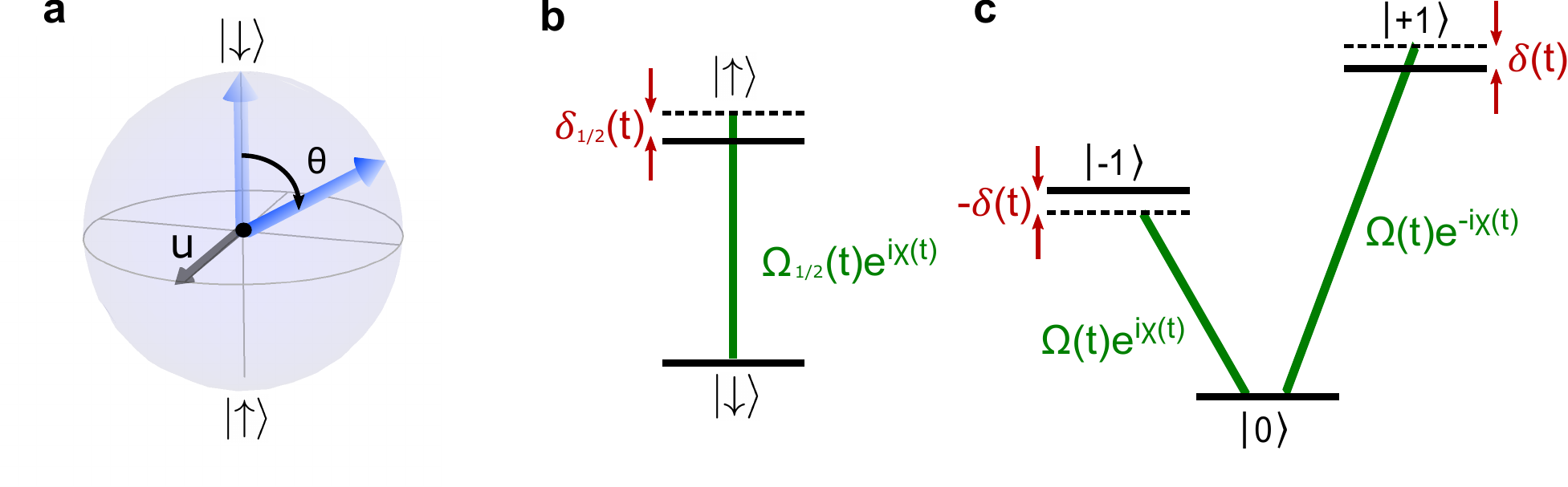}
\caption{{\bf Use of an effective two-level system to generate three-level control methods.} {\bf a}, We wish to implement a control method which transforms an initial state $\ket{\psi_{1/2}}_\mathrm{i}$ on the effective two-level Bloch sphere (which here we take to be $\ket{\downarrow}$), into a final state $\ket{\psi_{1/2}}_\mathrm{f} = e^{-i\theta\bold{\hat{u}}\cdot\bold{S}}\ket{\psi_{1/2}}_\mathrm{i}$, where $\bold{\hat{u}}$ is the axis of rotation and $\theta$ is the angle, (equivalent to $\ket{\psi_j}_\mathrm{f} = e^{-i\theta \bold{\hat{u}}\cdot\bold{J}}\ket{\psi_j}_\mathrm{i}$ in the real multi-level system). {\bf b}, In the effective two-level system, most control methods are implemented by applying a single control field of Rabi frequency $\Omega_\text{1/2}(t)$, instantaneous detuning $\delta_\text{1/2}(t)$ and phase $\chi(t)$. {\bf c}, By inverting the Majorana decomposition, we derive the control fields that we must apply to our real physical system, namely two fields of equal Rabi frequency $\Omega(t)$ and equal and opposite detunings and phases $\pm\delta(t)$ and $\pm\chi(t)$ respectively.} 
\label{fig:majorana}
\end{figure*}

\section{Introduction}

Quantum control methods are essential in many areas of experimental quantum physics, including trapped atoms, ions and molecules and solid state systems \cite{Glaser, Mabuchi, Vandersypen}. Although the focus is often on two-level systems, in nearly all experimental realisations a larger number of states need to be taken into consideration, for example to prepare a qubit in a two-level subspace of the system or to read out the state at the end of an experiment. In addition, the unique features of multi-level systems have led to new fields of research including electromagnetically induced transparency \cite{Fleischhauer} and single photon generation \cite{Kuhn}. Multi-level systems are also widely used in quantum computing, with applications such as the preparation and detection of dressed-state qubits \cite{Timoney, Webster}. A variety of multi-level methods including stimulated Raman adiabatic passage (STIRAP) \cite{Kuklinski}, multi-state extensions of Stark-chirped rapid adiabatic passage (SCRAP) \cite{Rangelov} and other adiabatic methods involving chirped laser fields \cite{Broers,Melinger,Vitanov} have been developed, in addition to numerical algorithms for optimised quantum control \cite{Khaneja}. However the development of new control methods for multi-level systems (especially for high-fidelity operations) is challenging and often inhibited by the mathematical complexity of such higher-dimensional Hilbert spaces. Previous investigations into multi-level dynamics have studied coherent excitation of multi-level systems under the action of SU(2) Hamiltonians \cite{Majorana, Bloch, Hioe, CookShore, Genov}. They showed that for a Hamiltonian with this symmetry there exists an equivalent Hamiltonian acting on a two-level system, and the dynamics of this two-level Hamiltonian can then be used to find solutions for the dynamics of the higher-dimensional system.

Here, we apply this insight to find states in a two-level system that are equivalent to the states we wish to transform between in the multi-level system. This is key for practical quantum control methods where it is often necessary to transfer population between two particular states with high fidelity \cite{Glaser}. If such states exist, any method to move between them can be transformed into the multi-level case. Thus, we can transform robust, high-fidelity two-level methods into new multi-level methods which also possess these desirable properties. We experimentally implement two novel control methods for trapped ions generated using the technique, demonstrating their high-fidelity and robustness. 

The manuscript is organized as follows. In section \ref{sec:Majorana}, we introduce the Majorana decomposition and detail how to design multi-level control methods using equivalent two-level methods. In section \ref{sec:threelevel} we introduce a three-level example system in $\yb$ and discuss the mapping to a two-level system for this specific case. In sections \ref{sec:adiabatic} and \ref{sec:composite}, we demonstrate adiabatic and composite control methods based on the Majorana decomposition in our trapped ion system. Finally, in section \ref{sec:fidelity}, we present a measurement of the fidelity of the two control methods.

\section{Majorana Decomposition}\label{sec:Majorana}

The Majorana decomposition was originally devised as a way of simplifying the dynamics of a spin-$j$ system in an inhomogeneous magnetic field, by reducing the dynamics to that of an effective two-level system \cite{Majorana, Bergeman, Bauch, Bloch}. Consider a Hamiltonian that takes the same form as a spin in a magnetic field, that is $H_j=\bold{\Lambda}(t)\cdot\bold{J}$ where $\bold{J} = J_x\hat{\bold{x}} + J_y\hat{\bold{y}} + J_z\hat{\bold{z}}$, $J_i$ being the angular momentum operators of a spin-$j$ particle, and $\bold{\Lambda}(t)$ is a three-component vector specifying the control fields that we apply to our system. Such a system can be said to have SU(2) symmetry \cite{CookShore}. Majorana showed that the dynamics of such a system can be mapped exactly onto the dynamics of a spin-1/2 particle, acted upon by the Hamiltonian $H_{1/2}=\bold{\Lambda}(t)\cdot\bold{S}$, $\bold{S}$ being the spin-1/2 angular momentum operator. This decomposition has been used to develop analytical solutions for the dynamics of a multi-level system \cite{CookShore, Hioe}. Here we apply these ideas to generate new high-fidelity multi-level quantum control methods. First we use the Majorana decomposition to transform a multi-level problem into its much simpler two-level equivalent, for which a multitude of control methods are readily available. By then inverting the Majorana decomposition, we obtain the control fields for a new equivalent multi-level method.

In order to describe this technique, we introduce the following mathematical framework, which expresses each step of the process in simple, geometrical terms. First, consider an initial and final state in a multi-level system which we require to be related by a rotation $\ket{\psi_j}_\mathrm{f} =e^{-i\theta \bold{\hat{u}}\cdot\bold{J}}\ket{\psi_j}_\mathrm{i}$, where $\bold{\hat{u}}$ and $\theta$ specify the axis and angle of rotation. The Majorana decomposition tells us that there will be an equivalent transformation in the spin-$1/2$ system: $\ket{\psi_{1/2}}_\mathrm{f} =e^{-i\theta \bold{\hat{u}}\cdot\bold{S}}\ket{\psi_{1/2}}_\mathrm{i}$ (Fig. \ref{fig:majorana}a), where the choice of $\ket{\psi_{1/2}}_\mathrm{i}$ is arbitrary. At this point we can use any of the many robust two-level control methods to carry out the transformation $\ket{\psi_{1/2}}_\mathrm{i} \to \ket{\psi_{1/2}}_\mathrm{f}$. To transform this two level method into the new multi-level control method we apply the inverse of the Majorana decomposition. Noting that any two-level Hamiltonian can be written in the form $H_{1/2}=\bold{\Lambda}(t)\cdot\bold{S}$, we obtain the multi-level method by producing a Hamiltonian $H_j$ with the same control vector $\bold{\Lambda}(t)$. This will perform the required multi-level state transformation $\ket{\psi_{j}}_\mathrm{i} \to \ket{\psi_{j}}_\mathrm{f}$. The new multi-level method will share desirable properties with the original two-level method, such as robustness to certain parameter errors that also have SU(2) symmetry. 

As an example, suppose that we want to transfer population between eigenstates of two different angular momentum operators in different directions. The initial and final states $\ket{\psi_j}_\mathrm{i}$ and $\ket{\psi_j}_\mathrm{f}$ are eigenstates of the projection angular momentum operators along the directions $\bold{\hat{r}}_i$ and $\bold{\hat{r}}_f$ respectively with the same eigenvalue $m_J$. Any rotation that transforms $\bold{\hat{r}}_i$ to $\bold{\hat{r}}_f$ will suffice. The simplest rotation (smallest rotation angle) is given by $\theta = \sin^{-1}(|\bold{r}_i\times\bold{r}_f|)$, $\bold{\hat{u}}=\bold{r}_i\times\bold{r}_f/|\bold{r}_i\times\bold{r}_f|$. For example, consider the $J_z$ and $J_x$ eigenstates for the j=1 three-level system. The $J_z$ eigenstates are the basis states \ket{+1}, \ket{0}, and \ket{-1}, with eigenvalues +1, 0, -1 respectively, while the three eigenstates of $J_x$ are $\ket{u}=\frac{1}{2}\ket{+1}+\frac{1}{2}\ket{-1}+\frac{1}{\sqrt{2}}\ket{0}$, $\ket{D}=(\ket{+1}-\ket{-1})/\sqrt{2}$, and $\ket{d}=\frac{1}{2}\ket{+1}+\frac{1}{2}\ket{-1}-\frac{1}{\sqrt{2}}\ket{0}$, again with eigenvalues of +1, 0,  and -1. We can consider the effect of consecutive rotations of $\pi/2$ about the $y$ axis, that's to say applications of the rotation operator $e^{-i(\pi/2)J_y}$. If we start in the state \ket{0}, then ignoring global phases we get the following sequence of states:
\begin{equation}\label{eq:0D}
\ket{0}\rightarrow\ket{D}\rightarrow\ket{0}\rightarrow\ket{D}\rightarrow\ket{0}
\end{equation}
where the ion is alternating between the $m=0$ eigenstates of the two angular momentum operators $J_z$ and $J_x$, since the $m_J=0$ eigenstates of a projection operator and its inverse are equal. If instead we start in $\ket{+1}$ we get the sequence
\begin{equation}\label{eq:pum}
\ket{+1}\rightarrow\ket{u}\rightarrow\ket{-1}\rightarrow\ket{d}\rightarrow\ket{+1}, 
\end{equation}
where the ion is moving between the $\pm1$ eigenstates of the $J_z$ and $J_x$ operators. Any two-level control method that rotates by an angle $\pi/2$ about the $y$ axis can therefore be used to transform between states in the three-level system linked in equations \ref{eq:0D} and \ref{eq:pum}.

The method described in this section is a general technique to derive new robust quantum control methods for multi-level systems based on the Majorana decomposition. In the following sections we will describe a specific physical system of interest, a three-level system in the ground state of a single trapped ion, and demonstrate the application of this method to robustly perform a specific desired state transformation within this system.

\section{Three-level Trapped Ion System}\label{sec:threelevel}

To illustrate the technique described in section \ref{sec:Majorana}, we generate new control methods for the coherent manipulation of a three-level V-system. We demonstrate these methods experimentally using a single trapped $^{171}$Yb$^+$ ion, where the three levels \ket{+1}, \ket{0} and \ket{-1} correspond to the states \ket{F=1, m_F=+1}, \ket{F=0} and \ket{F=1, m_F=-1} of the $^2S_{1/2}$ ground-state hyperfine manifold respectively. The ion is confined in a linear Paul trap, details of which are described in Refs. \cite{McLoughlin2, Lake}. A magnetic field of $B_0 = 8.8305(4)\,$G is applied using permanent magnets inside the vacuum system and external current coils. The magnetic field splits the energies of the states making up the $F=1$ manifold. The transitions from $\ket{0}$ to $\ket{\pm1}$ are driven by two microwave fields generated using an RF arbitrary waveform generator, which creates a waveform with a bandwidth of $\approx 30\,$MHz centred around $100\,$MHz. Typically we set the Rabi frequencies $\Omega_1$ and $\Omega_2$ of these applied fields to be equal, so that the dark state $\ket{D}$ will be an eigenstate of the dressed Hamiltonian (Eq. \ref{eq:majoranaH}). The waveform is then frequency mixed with a signal near $12.5\,$GHz, before being amplified to $2\,$W and sent to a microwave horn positioned near a viewport of the vacuum system, approximately $2\,$cm from the ion. The ion is prepared in $\ket{0}$ using optical pumping and a fluorescence measurement distinguishes between $\ket{0}$ and $\{\ket{-1},\ket{0'},\ket{+1}\}$, where $\ket{0'} \equiv \ket{F=1, m_F = 0}$ is an additional state in the $F=1$ manifold that is not used. A maximum likelihood method is used to normalise the data against independently measured state detection errors (Appendix \ref{sec:statistic}).

\begin{figure*}
\centering
\includegraphics[width=1.7\columnwidth]{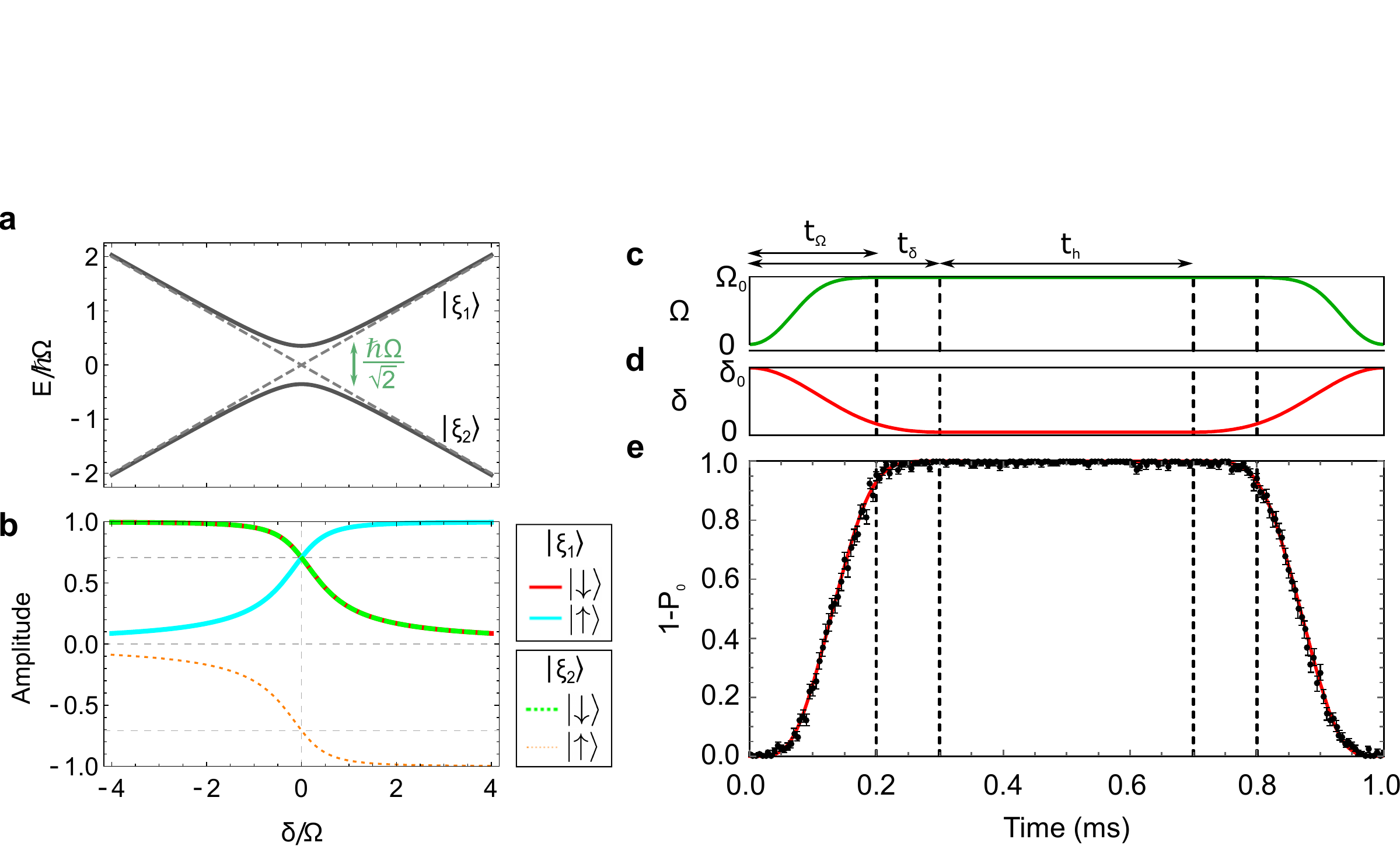}
\caption{{\bf Adiabatic transfer to the dark state of a dressed three-level system}. {\bf a}, Energy eigenvalues and {\bf b,} eigenstates $\{\ket{\xi_1},\ket{\xi_2}\}$ of $H_{1/2}$ as a function of $\delta/\Omega$ for $\chi = 0$. {\bf b}, Shows analytically calculated amplitudes of these eigenstates, all of which can be defined as real numbers in this case. An avoided crossing is present at $\delta/\Omega = 0$, at which point the eigenstates are the dressed states $(\ket{\downarrow} \pm \ket{\uparrow})/\sqrt{2}$, which are separated in energy by $\hbar\Omega/\sqrt{2}$. Therefore, by adiabatically varying the detuning and Rabi frequency, the population can be coherently transferred from $\{\ket{\downarrow},\ket{\uparrow}\}$ to $(\ket{\downarrow} \pm \ket{\uparrow})/\sqrt{2}$. {\bf c-e}, Demonstrating the method using a single $\yb$ ion. In the three-level system, the adiabatic procedure will transfer population from $\ket{0}$ to the dark state $\ket{D} = (\ket{+1} - \ket{-1})/\sqrt{2}$.  {\bf c-d}, The temporal profiles for the Rabi frequency $\Omega$ (solid green line in {\bf c}) and the instantaneous detuning $\delta$ (solid red line in {\bf d}), where the relevant parameters are given in the text. {\bf e}, Measured probability for the ion to be in the $\yb$ $F = 1$ state given by $P(F=1) = 1 - P_0$ as a function of time. Each point is the average of 300 repetitions. The theoretical probability for the ion to be in $F = 1$ as a function of time (solid red line) is obtained from a numerical simulation of the system with no free parameters, which can be seen to agree well with the measured data.}
\label{fig:singlemap}
\end{figure*}

We would like to transfer the system from \ket{0} to the superposition state $\ket{D}\equiv(\ket{+1}-\ket{-1})/\sqrt{2}$, which can be protected against decoherence caused by fluctuating magnetic fields by the application of a pair of dressing fields \cite{Timoney, Webster} and has been shown to be useful for quantum computation \cite{Timoney, Webster, Randall, Weidt2,Weidt,Bermudez3} and magnetometry \cite{Baumgart}. Previous methods to transfer population between these states are either susceptible to errors from fluctuating magnetic fields \cite{Timoney, Webster} or require the use of the $\ket{0'}$ state, which would ideally be reserved to form a qubit along with $\ket{D}$ \cite{Randall}. It would therefore be desirable to design a robust method to transfer between these states with low infidelity. The required population transfer corresponds to the unitary transformation $U_{j=1}=e^{-i(\pi/2)J_y}$, a rotation about the $y$-axis by $\pi/2$. Due to the Majorana decomposition, this is equivalent to the transformation $\ket{\downarrow}\rightarrow\frac{1}{\sqrt{2}}(\ket{\downarrow}+\ket{\uparrow})$ in a spin-1/2 system, as shown in section \ref{sec:Majorana} (Fig. \ref{fig:majorana}a).

There are many ways to carry out this two-level process, such as a simple $\pi/2$ pulse, or more robust methods such as composite pulses and adiabatic passage. The vast majority of two-level methods that can be implemented use a single control field, with possibly time-varying amplitude, frequency and phase (Fig. \ref{fig:majorana}b). Moving to an interaction picture rotating at the frequency of the field, and after making the rotating wave approximation, this corresponds to a Hamiltonian
\begin{equation}\label{eq:H1/2}
H_\text{1/2} = \frac{\hbar}{2}\left(\begin{array}{cc}
-\delta_\text{1/2}(t) & \Omega_\text{1/2}(t) e^{i\chi(t)} \\ 
\Omega_\text{1/2}(t) e^{-i\chi(t)} & \delta_\text{1/2}(t)
\end{array} \right)
\end{equation}
(with the states ordered $\ket{\downarrow}, \ket{\uparrow}$), which can be written as $H_{1/2} = \hbar(\Omega_\text{1/2}(t)\cos(\chi(t))S_x+\Omega_\text{1/2}(t)\sin(\chi(t))S_y+\delta_\text{1/2}(t)S_z)$, where $\Omega_\text{1/2}(t)$, $\delta_\text{1/2}(t)$ and $\chi(t)$ are the time varying Rabi frequency, instantaneous detuning, and phase, respectively. Once the forms of $\Omega_\text{1/2}(t)$, $\delta_\text{1/2}(t)$ and $\chi(t)$ have been chosen to perform the required transformation $\ket{\downarrow}\rightarrow\frac{1}{\sqrt{2}}(\ket{\downarrow}+\ket{\uparrow})$, we can invert the Majorana decomposition to determine what real-world control fields we must apply to our physical three-level system to move between the initial and final states $\ket{0}$ and $\ket{D}$. The resulting three-level Hamiltonian is obtained by replacing the Pauli matrices in $H_{1/2}$ above with the three-level spin matrices $J_i$: $H_{j=1} = \hbar(\Omega_{1/2}(t)\cos(\chi(t))J_x+\Omega_{1/2}(t)\sin(\chi(t))J_y+\delta_{1/2}(t)J_z)$. This Hamiltonian can be written as
\begin{equation}\label{eq:majoranaH}
H_{j=1} = \frac{\hbar}{2} \left(\begin{array}{ccc}
-\delta(t) & \Omega(t) e^{i\chi(t)} & 0 \\ 
\Omega(t) e^{-i\chi(t)} & 0 & \Omega(t) e^{i\chi(t)} \\ 
0 & \Omega(t) e^{-i\chi(t)} & \delta(t)
\end{array} \right)
\end{equation}
(with the states ordered $\ket{-1}, \ket{0}, \ket{+1}$), which corresponds to a pair of control fields, each of Rabi frequency $\Omega(t) = \sqrt{2}\Omega_\text{1/2}(t)$, with opposite phases $\pm\chi(t)$ and opposite detunings $\pm\delta(t) = \pm2\delta_\text{1/2}(t)$ (Fig. \ref{fig:majorana}c).

Now that we have derived a transformation between the effective two-level system and our physical three-level system, we can design new control methods to achieve the desired mapping based on existing two-level control methods. Quantum control methods for two-level systems are often designed to protect against errors caused by fluctuating parameters, such as detuning and Rabi frequency. These errors in a two-level system will also have equivalents in the multi-level case, and any protection offered will carry over. In the $\yb$ system used here, two main sources of error are caused by magnetic field noise and common mode Rabi frequency noise, the effects of which both have SU(2) symmetry and can therefore be countered by the appropriate choice of two-level control method. In the following sections, we design and demonstrate two such methods.

\section{Adiabatic Control Method}\label{sec:adiabatic}

The first method is an adiabatic method following on from the work of Hioe \cite{Hioe}, which is the three-level equivalent of the well-known two level process of rapid adiabatic passage described by the Landau-Zener-Stuckelberg-Majorana model \cite{Landau, Zener}. Here, population is transferred between two states by adiabatically moving their energies to an avoided crossing. If the field is adiabatically varied from the regime where $\Omega_{1/2}/\delta_{1/2} = 0$, to $\Omega_{1/2}/\delta_{1/2} = \infty$ with $\chi = 0$ by turning the field on slowly whilst chirping the frequency, the population will be transferred from the initial eigenstate $\ket{\downarrow}$ to $(\ket{\downarrow} + \ket{\uparrow})/\sqrt{2}$ (see Fig. \ref{fig:singlemap}a,b). If we translate this into the three-level picture, we obtain a Hamiltonian of the form shown in equation \ref{eq:majoranaH}. This describes a novel adiabatic process involving chirped pulses and amplitude shaping which transfers population from $\ket{0}$ to $\ket{D}$, similar to the analytical solution derived by Hioe \cite{Hioe}.

A Blackman function \cite{Blackman} is used to define the form of the time-varying detuning $\delta(t)$. This pulse shape was chosen because in numerical simulations it produced the lowest infidelity due to non-adiabaticities. For a Blackman chirp profile starting at $\delta_0$ and finishing at zero detuning, the required `instantaneous' detuning is
\begin{equation}
\delta(t) = \frac{\delta_0}{50}\left( 21 + 25\cos\left(\frac{\pi t}{t_\delta}\right) + 4\cos\left(\frac{2\pi t}{t_\delta}\right)\right),
\end{equation}
where $t_\delta$ is the detuning chirp time (Fig. \ref{fig:singlemap},d). Due to the choice of interaction picture chosen to derive equations \ref{eq:H1/2} and \ref{eq:majoranaH}, where the interaction frame is rotating at the time-dependent frequency of the field, this is the detuning used in these equations. In the lab frame, the required frequency of the physical field is given by $\omega_0 + \Delta(t)$, where $\omega_0$ is the resonant frequency and $\Delta(t)$ is the detuning. However, $\Delta(t)$ is not equal to this instantaneous detuning of the field. The instantaneous frequency of a sinusoidal function at any given time is given by the time derivative of its overall phase, which in our case is equal to $\delta(t) = d(\Delta(t)t)/dt$ for $\chi(t) = 0$. The required profile for $\Delta(t)$ is therefore given by
\begin{equation}\label{eq:chirp}\begin{split}
	\Delta(t) &= \frac{1}{t}\int_0^{t} \delta(\tau) d\tau \\
	&= \frac{\delta_0}{50 t}\left[21t + \frac{t_\delta}{\pi}\left(25\sin\left(\frac{\pi t}{t_\delta}\right) + 2\sin\left(\frac{2\pi t}{t_\delta}\right)\right)\right].
\end{split}\end{equation}
The amplitude of the driving fields are also changed during the first part of the detuning chirp. We again use a Blackman function, giving a Rabi frequency profile 
\begin{equation}
\Omega(t) = \frac{\Omega_0}{50}\left( 29 - 25\cos\left(\frac{\pi t}{t_\Omega}\right) - 4\cos\left(\frac{2\pi t}{t_\Omega}\right)\right),
\end{equation}
where $t_\Omega$ is the amplitude ramp time (Fig \ref{fig:singlemap},c). The Rabi frequency is then kept constant at $\Omega_0$ until the detuning chirp is complete.

We implement this procedure experimentally in our $\yb$ ion system. Fig. \ref{fig:singlemap}e shows the probability of measuring the system in the $\yb$ $F=1$ level ($1 - P_0$) as a function of time during the adiabatic procedure. First, the transformation \ket{0}$\rightarrow$\ket{D} is performed. Next, the system is left in the state $\ket{D}$, which is protected by the control fields, for a `hold' time $t_h = 400\,\mu$s. Finally, the inverse transformation \ket{D}$\rightarrow$\ket{0} is performed by reversing the amplitude shaping and chirped frequency profiles of the forward process. The optimal parameters for the Blackman profiles were found by simulations to be $\Omega_0/2\pi = 40\,$kHz, $\delta_0/2\pi = 60\,$kHz, $t_\Omega = 200\,\mathrm{\mu s}$ and $t_\delta = 300\,\mathrm{\mu s}$. Compression in the microwave amplifiers slightly alters the amplitude envelope of the applied microwave radiation compared with that generated by the arbitrary waveform generator. This effect, which has been included in the numerical simulation, has a negligible impact on the simulated fidelity. Plots of $\Omega(t)$ and $\delta(t)$ in Fig. \ref{fig:singlemap}c,d, include these effects of compression.

\begin{figure*}
\centering
\includegraphics[width=2\columnwidth]{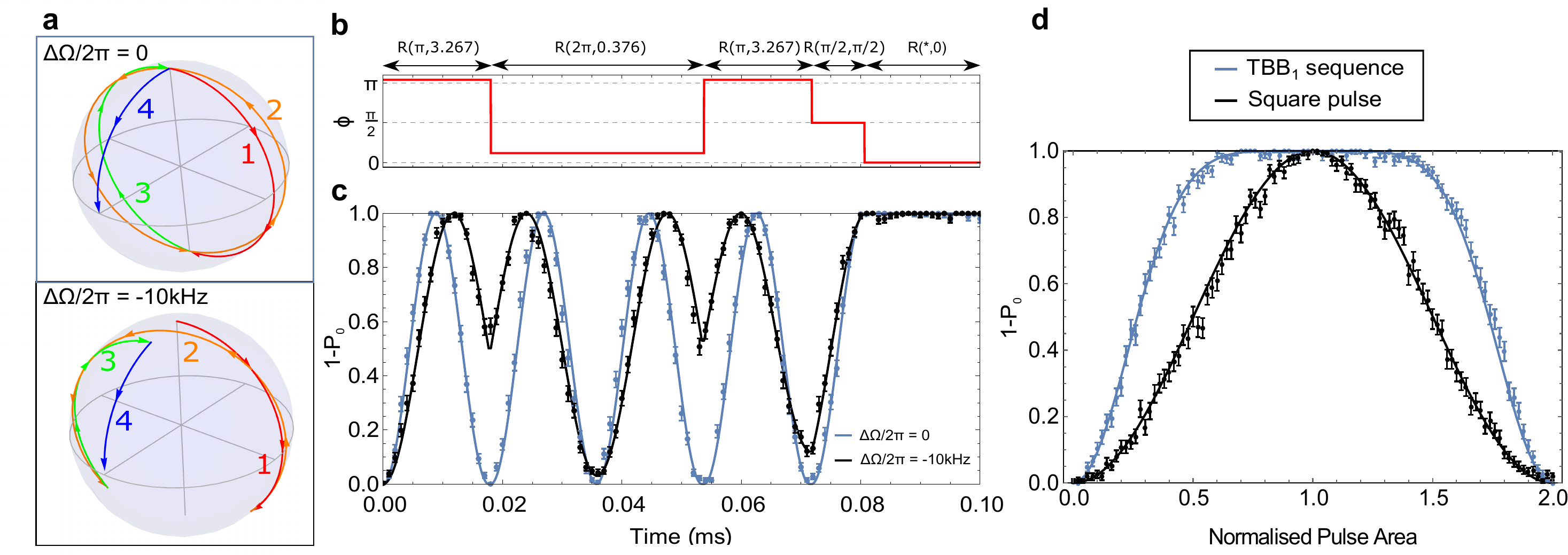}
\caption{{\bf Robust population transfer to the dark state using the $\text{TBB}_1$ composite pulse sequence}. {\bf a}, The $\text{TBB}_1$ composite pulse sequence represented on the effective two-level Bloch sphere. The sequence consists of four resonant pulses with varying pulse area and phase which can be written as a sequence of rotations on the Bloch sphere of the form $R(\theta_R, \phi_R)$. Each of these rotations is represented as a coloured line on the Bloch sphere, in the order red, orange, green, blue (numbered to show ordering). Above is the trajectory in the case of zero Rabi frequency error and below for the $\Delta\Omega = -2\pi \times 10\,$kHz case. We implement both of these cases experimentally to demostrate the robustness of the method to pulse area errors. {\bf b}, The phase $\chi$ as a function of time implementing the $\text{TBB}_1$ pulse sequence for a fixed Rabi frequency $\Omega_0/2\pi = 40\,$kHz. An extra phase change of $-\pi/2$ at the end ensures the population remains in $\ket{D}$ after the procedure. Therefore the total pulse sequence is $R(*,0)\cdot R(\pi/2,\pi/2)\cdot R(\pi,3.267)\cdot R(2\pi,0.376)\cdot R(\pi,3.267)$. {\bf c}, The measured population in F=1 as a function of time with $\Delta\Omega = 0$ ({\bf a} upper sphere, {\bf c} light blue line (light grey when printed in greyscale)) and for a Rabi frequency error of $\Delta\Omega = -2\pi \times 10\,$kHz ({\bf a} lower sphere, {\bf c} black line), showing that the sequence is robust to such errors. {\bf d}, Measured population in F=1 as a function of pulse area for a square $\pi/2$ pulse (black) and the new $\text{TBB}_1$ pulse sequence (light blue (light grey when printed in greyscale)), demonstrating that the $\text{TBB}_1$ sequence maintains the robustness to pulse area error of the original two-level $\text{BB}_1$ sequence. The pulse area is normalised such that the nominal pulse area for a $\pi/2$ rotation is 1. The solid lines in {\bf c} and {\bf d} correspond to numerical simulations of the sequence with no free parameters.}
\label{fig:bb1}
\end{figure*}

\section{Composite Control Method}\label{sec:composite}

We have shown that our technique can be used to develop a three-level adiabatic method similar to the two-level method of rapid adiabatic passage. As a further demonstration of our technique to develop novel multi-level control methods, we implement a resonant control method to transfer population from $\ket{0}$ to $\ket{D}$. We do this by creating a three-level composite pulse sequence. A widely used example of a two-level composite pulse sequence is the $\text{BB}_1$ pulse sequence by Wimperis \cite{Wimperis}, which consists of four resonant Rabi pulses and can protect against pulse area errors. The four pulses of the $\text{BB}_1$ sequence carry out four consecutive rotations of the type $R(\theta_R,\phi_R)$ each with a particular choice of rotation angle $\theta_R$ and phase $\phi_R$ (corresponding to a rotation axis $\hat{\bold{u}}=\cos{\phi_R}\hat{\bold{x}} + \sin{\phi_R}\hat{\bold{y}}$). For a rotation from $\theta = \phi = 0$ to $\theta = \pi/2$, $\phi = 0$, it consists of four pulses and is given by $U(\text{BB}_1) = R(\pi/2,\pi/2)\cdot R(\pi,3.267)\cdot R(2\pi,0.376)\cdot R(\pi,3.267)$, where $R(\theta_R,\phi_R)$ is a rotation on the Bloch sphere by polar angles $\theta_R$ and $\phi_R$ (see Fig. \ref{fig:bb1}a). 

Using our technique, we can produce an analogous control method for three-level systems which can robustly transfer population from $\ket{0}$ to $\ket{D}$ (which we call the $\text{TBB}_1$ sequence). This method consists of a sequence of simultaneous microwave pulses on the $\ket{0}$ to $\ket{\pm1}$ transitions, with parameters set such that $\Omega_0 t/\sqrt{2} = \theta_R$, $\chi = \phi_R$ and $\delta = 0$. Therefore the three-level $\text{TBB}_1$ sequence consists of four pulses of length $17.7, 35.4, 17.7$ and $8.8 \mathrm{\mu s}$ and phases $\pm 1.63, \pm 0.19, \pm 1.63$ and $\pm 0.79$ on the $\ket{0}$ to $\ket{\pm 1}$ transitions. Thus a rotation from $\ket{0}$ to $\ket{D}$ (which again is $\ket{\downarrow}$ to $(\ket{\downarrow} + \ket{\uparrow})/\sqrt{2}$ in the effective two-level system) is implemented. In order to protect the $\ket{D}$ state after the sequence, the control fields are simply left on, with the relative phase $\chi$ set to 0. Figure \ref{fig:bb1}c shows the population in $F = 1$ as a function of time during the $\text{TBB}_1$ pulse sequence for two cases. In one case the Rabi frequency is set to the correct value such that $\Delta\Omega = \Omega - \Omega_0 = 0$, while in the second case the Rabi frequency is deliberately mis-set by $\Delta\Omega = -2\pi \times 10\,$kHz, which corresponds to a $25\%$ error in the applied microwave amplitude. It can be seen that in both cases the final population is almost entirely transferred to the $F=1$ manifold, demonstrating the robustness of the composite sequence to substantial errors in the pulse area. The $\text{TBB}_1$ sequence is completed in a time of $80\,\mu$s compared to $300\,\mu$s for the adiabatic method, but both methods could be sped up by increasing the applied microwave power (i.e. raising the Rabi frequency). Figure \ref{fig:bb1}d shows the population in $F =1 $ as a function of normalised pulse area for a single pulse nominally driving a rotation $R(\pi/2,\pi/2)$ in the effective two-level system, as well as when the $\text{TBB}_1$ pulse sequence is applied. The improvement in robustness of the $\text{TBB}_1$ sequence compared to the single pulse can clearly be seen, demonstrating that composite quantum control techniques developed for two-level systems give the same advantages in the three-level case.

\begin{figure}
\centering
\includegraphics[width=\columnwidth]{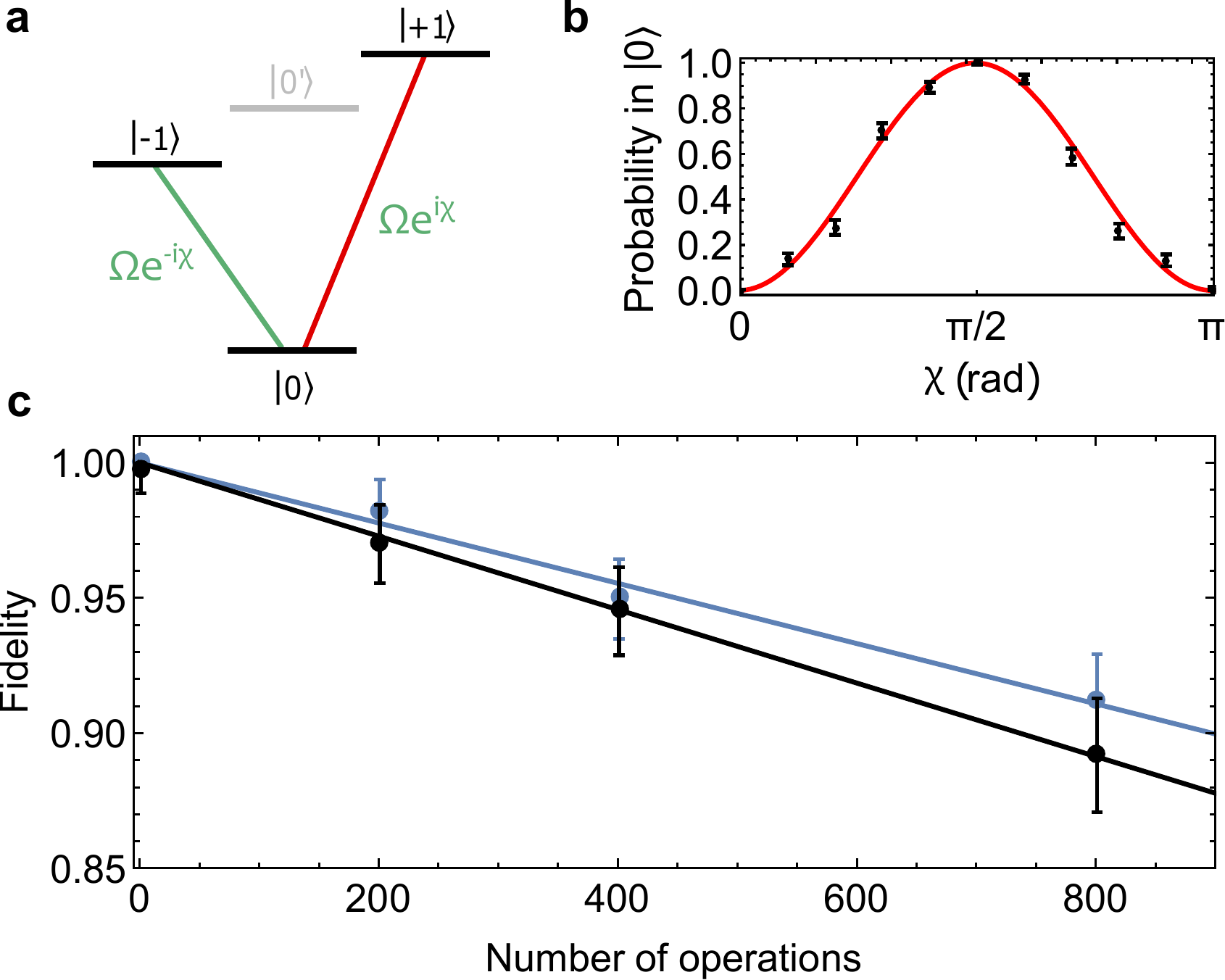}
\caption{{\bf Measuring the fidelity.} {\bf a}, The fidelity with which we produce the $\ket{D}$ state can be obtained by applying two fields resonant with the $\ket{0} \leftrightarrow \ket{+1}$ and $\ket{0} \leftrightarrow \ket{-1}$ transitions with equal Rabi frequency $\Omega$, and varying the phase $\chi$ of the two fields in equal and opposite directions. {\bf b}, The measured population in $\ket{0}$ as a function of $\chi$ after a single adiabatic operation (black points), which can be fitted to the function $A_0 + A\cos(2\chi + \phi_0)$ (solid red curve) to extract the mapping fidelity using a maximum likelihood fitting method (Appendix \ref{sec:statistic}). Each point is the average of 200 repetitions. {\bf c}, The fidelity as a function of the number of applications of the adiabatic method (black) and resonant $\text{TBB}_1$ sequence (light blue (light grey when printed in greyscale)). For the adiabatic method, the population transfer back to $\ket{0}$ begins immediately after it reaches $\ket{D}$ ($t_h = 0$). A linear least squares fit to the data gives an average infidelity per operation of $1.4(4) \times 10^{-4}$ for the adiabatic method and $1.1(4) \times 10^{-4}$ for the composite pulse sequence.}
\label{fig:mapfidelity}
\end{figure}

\section{Fidelity Measurements}\label{sec:fidelity}

Although the data presented in Figs. \ref{fig:singlemap} and \ref{fig:bb1} show a good agreement with theory, the florescence measurement scheme used can only determine the values of the quantities $P_0$ and $P_{+1} + P_{-1} + P_{0'} = 1 - P_0$, where $P_j = \bra{j}\rho\ket{j}$ and $\rho$ is the measured state. This is not sufficient to calculate the fidelity with which the state $\ket{D}$ is prepared. Therefore a more complex method is required to fully characterise the state fidelity. The fidelity of $\ket{D}$ is given by 

\begin{equation}\label{eq:fid}
\mathcal{F}_D \equiv \bra{D}\rho\ket{D} = \frac{1}{2}(P_{+1} + P_{-1}) + |\rho_{\pm}|\cos( \phi_{\pm}),
\end{equation} 
where we have written the off-diagonal matrix elements in polar form as $\rho_{+1,-1} \equiv |\rho_{\pm}|e^{i \phi_{\pm}} = \rho_{-1,+1}^*$. To measure this fidelity, an additional resonant pulse on the $\ket{0}$ to $\ket{\pm1}$ transitions (Eq. \ref{eq:majoranaH}, Fig. \ref{fig:mapfidelity}a) is applied for a time $t = \pi/2\Omega_{1/2}$ (we apply this pulse simply by leaving the microwave fields on after the sequence and stepping the phase by $\chi$). If the phase $\chi$ is varied, the population in $\ket{0}$ is given by
\begin{equation}
P_0(\chi) = \frac{1}{2}(P_{+1} + P_{-1}) + |\rho_{\pm}|\cos(2\chi + \phi_{\pm}),
\end{equation}
where $P_{+1}$, $P_{-1}$, and $\rho_{\pm}$ are density matrix elements of the state before the additional pulse is applied. Comparing with Eq. \ref{eq:fid}, it can be seen that the offset, amplitude and phase offset of the resulting sinusoidal curve can be used to calculate $\mathcal{F}_D$. Fig. \ref{fig:mapfidelity}b shows the result of such an experiment after a single adiabatic transfer operation from $\ket{0}$ to $\ket{D}$. The data is fitted using maximum likelihood estimation (Appendix \ref{sec:statistic}) with the fit function $A_0 + A\cos(2\chi + \phi_0)$, giving fit parameters $A_0 = 0.500(4)$, $A = 0.500(3)$ and $\phi_0 = 3.16(3)$. This gives a map infidelity of $\mathcal{F}_D = 1.000(7)$. To obtain a more accurate infidelity estimate we must average over a large number of operations. The fidelity can be measured after $N$ operations for multiple values of $N$, from which the average infidelity $1-\overline{\mathcal{F}}_D$ can be calculated. This method is used to calculate the average fidelities of both the adiabatic and composite quantum control procedures. We measure an average infidelity per operation of $1.4(4) \times 10^{-4}$ for the adiabatic method and $1.1(4) \times 10^{-4}$ for the composite pulse sequence \footnote{A recent study has suggested that the true infidelities may actually be somewhat lower than this, as infidelities due to coherent effects may be overestimated by this method \cite{Vitanov3}. However, as we believe our infidelities to be dominated by incoherent effects such as decoherence, this effect may not be significant.}.

The experimentally achieved fidelity of the adiabatic control method is determined by two factors: the first is infidelities introduced during the operation due to non-adiabaticity of the frequency and amplitude modulation and decoherence, and the second is the precision with which the parameters of the applied radiation fields can be set, as they determine the final state obtained, which we call $\ket{\psi_\text{dr}}$. By repeatedly applying the forward and reverse adiabatic operations we can determine the first of these infidelities, as to first order they will be amplified by the number of repeats to a measurable level. We do not attempt to measure the second infidelity $1 - |\bracket{D}{\psi_\text{dr}}|^2$ as we do not have a process to amplify this infidelity, and any direct measurement is subject to the same inaccuracies in parameter setting. Instead we can estimate the size of this infidelity given the precision we can set the parameters of the radiation fields. The parameters in question are how equal the Rabi frequencies of the two fields can be set, and the accuracy to which the detuning of the two radiation fields can be set to zero. We determined that we set the fractional accuracy of the Rabi frequencies $|\Omega_1-\Omega_2|/(\Omega_1+\Omega_2) < 0.0015$ and that each of the detunings are set such that $|\delta| < 3\,$Hz. From simulations, this leads to an infidelity of preparing $\ket{D}$ of $< 10^{-4}$. We also note that for many applications, such as the use of the $\ket{D}$ and $\ket{0'}$ states as a qubit, this second infidelity only has a small effect on the overall fidelity of qubit operations. This `dressed-state qubit' is used because the coherence of the qubit is protected against magnetic field fluctuations \cite{Timoney, Webster}. In the event of a slight Rabi frequency mismatch or detuning error, the dressed state produced will not be exactly $\ket{D}$, but this state and $\ket{0'}$ will still form a valid qubit which will still be insensitive to magnetic field noise to first order.

The measured infidelities are consistent with the lifetime of the $\ket{D}$ state, which was measured in a separate experiment to be $2.6(4)\,$s. The lifetime of $\ket{D}$ is limited by ambient magnetic field noise with frequency close to the dressed-state energy splitting. Since ambient noise generally scales as $\sim 1/f$, increasing the dressing field Rabi frequency is expected to improve this result \cite{Webster}. We have also verified that the coherence of a $\{\ket{0'},\ket{D}\}$ qubit is preserved throughout such an adiabatic transfer (Appendix \ref{sec:dressedqubit}).

\section{Conclusion}\label{sec:conclusion}

In this article, we have used the Majorana decomposition to develop a technique for generating new coherent control methods to transform between two desired multi-level states, based on existing two-level methods. This allows insights gained into robust control of two-level systems to be harnessed and applied to multi-level quantum control in a rigorous and analytical way. We have applied this technique to two well known composite pulse and adiabatic methods to create new three-level methods and have implemented these experimentally with high fidelity. These methods may be particularly important for the implementation of scalable quantum computing \cite{Lekitsch}. The technique we use to generate quantum control methods is general and can be applied to different quantum systems with arbitrary numbers of levels (Appendix \ref{sec:dlevels}). Furthermore, we have shown that the control methods generated can be robust and applied with high-fidelity. Therefore we believe this approach shows great promise for high-fidelity quantum control across a broad range of physical systems.   

\section*{Acknowledgements}
We would like to thank Bruce Shore for providing us with very useful insights on the theory of multi-level quantum control. This work is supported by the U.K. Engineering and Physical Sciences Research Council [EP/G007276/1; the U.K. Quantum Technology hub for Networked Quantum Information Technologies (EP/M013243/1) and the U.K. Quantum Technology hub for Sensors and Metrology (EP/M013294/1)], the European Commissions Seventh Framework Programme (FP7/2007-2013) under grant agreement no. 270843 Integrated Quantum Information Technology (iQIT), the Army Research Laboratory under cooperative agreement no. W911NF-12-2-0072, the U.S. Army Research Office under contract no. W911NF-14-2-0106, and the University of Sussex. NVV acknowledges support by the Bulgarian Science Fund Grant No. DN 18/14.

\appendix

\vline

\vline

\section{Statistical methods}\label{sec:statistic}

To normalise the data against state detection errors, before each experiment a histogram of fluorescence measurements is taken after preparing the ion in both the $\ket{0}$ and $\ket{0'}$ states, corresponding to dark and bright expected results respectively. Using a threshold of 2 photons, the detection fidelity is typically measured to be around $97\%$. A linear map can then be extracted from the measured errors, which gives the probability to measure a bright event as $p_b(p) = P(b|1)p + P(b|0)(1-p)$, where $p$ is the probability that the population was in the $F=1$ manifold and $P(b|1)$ and $P(b|0)$ are the probabilities for a bright measurement given that the ion was in the $F=1$ and $F=0$ manifolds, respectively. The data is scaled using a maximum likelihood method based on a binomial distribution. This maximises the log-likelihood function for a beta probability density function, given by

\begin{equation}
f_B = \sum_{i = 1}^N \log\left(\frac{(n+1)n! p_b(p_i)^{k_i} (1-p_b(p_i))^{n-k_i}}{k_i!(n-k_i)!}\right),
\end{equation}
where $n$ is the number of repetitions per data point, $N$ is the number of data points and $k_i$ is the number of bright events for the $i$th data point. For individual data points, $N = 1$ and therefore $p_1$ is found by maximising $f_B$ for $k_1$. To fit the fidelity measurements shown in Fig. \ref{fig:mapfidelity}, the probabilities are replaced by a fit function $p_i = A_0 + A\cos(2\chi_i + \phi_0)$. In this case, $f_B$ is maximised over all $N$ data points for different $\chi_i$, and the best fit parameters for $A_0$, $A$ and $\phi_0$ are extracted. The state fidelity is then given by $\mathcal{F}_D = A_0 - A\cos(\phi_0)$, which is plotted as a function of the number of maps in Fig \ref{fig:mapfidelity}c. A linear least squares fit is then applied with the fit function $1- x\epsilon_m$, where $x$ is the number of maps and $\epsilon_m = 1-\overline{\mathcal{F}}_D$ is the average infidelity per map.

\section{Spin-$j$ representation of arbitrary spin-1/2 unitaries}\label{sec:spinj}

As well as the mapping between initial and final states, it is also useful to derive a theoretical solution for the intermediate state of the multi-level system during application of the control fields. One option is to consider at arbitrary times during the transformation the equivalent rotation matrix in the multi-level system. However rather than doing this explicitly, the unitary operation in the multi-level system can be directly calculated from the unitary operation in the two-level system. The spin-1/2 state $\ket{\Psi_{1/2}} = a\ket{\downarrow}+b\ket{\uparrow}$ is obtained by applying the general unitary  
\begin{equation}\label{eq:U2}
U^\frac{1}{2} = \left(\begin{array}{cc}
a & -b^* \\ 
b & a^*
\end{array} \right)
\end{equation}
to the initial state $\ket{\downarrow}$. From this unitary, the unitary in the multi-level system can be calculated directly. For the general spin-$j$ system, the matrix elements of $U_{j}$ are given by \cite{Bloch,Torosov}
\begin{equation}\label{eq:Ud}\begin{split}
U_{rs}^j = &\sum_{q = q_\text{min}}^{q_\text{max}}\sqrt{C_q^{r-1} C_q^{s-1} C_{s-1-q}^{2j+1-r} C_{r-1-q}^{2j+1-s}} \\
&\times a^{2j+2-r-s+q}(a^*)^q b^{s-1+q} (-b^*)^{r-1+q},
\end{split}\end{equation}
where $q_\text{min} = \max[0,r+s-2j]$ and $q_\text{max} = \min[r-1,s-1]$ and $C_k^n = n!/[k!(n-k)!]$ is the binomial coefficient. For the $j=1$ case, this results in the unitary transformation \cite{Hioe}
\begin{equation}\label{eq:U}
U^{j=1} = \left(\begin{array}{ccc}
a^2 & - a b^* \sqrt{2} & b^{*2} \\ 
ab\sqrt{2} & |a|^2 - |b|^2 & -a^*b^*\sqrt{2} \\ 
b^2 & a^*b\sqrt{2} & a^{*2}
\end{array} \right).
\end{equation}

\section{Dressed state qubit mapping}\label{sec:dressedqubit}

In the context of a scalable microwave-driven trapped ion quantum computing architecture \cite{Lekitsch, Weidt}, it is useful to map the state of a qubit stored in the $\{\ket{0},\ket{0'}\}$ basis of an $\yb$ ion to the $\{\ket{D},\ket{0'}\}$ basis. This can be done by implementing either the adiabatic or the resonant method to transfer any population in state $\ket{0}$ to $\ket{D}$. While we have verified that this population transfer process can be implemented with high fidelity, this does not necessarily indicate that the coherence of the qubit is maintained throughout the population transfer process. Therefore we carried out a Ramsey-type experiment to measure the coherence of the qubit before and after the mapping, in the case of the adiabatic transfer method.

In these Ramsey experiments, we start with a resonant $\pi/2$ pulse on the $\ket{0}$ to $\ket{0'}$ `clock' transition to put the ion in the state $(\ket{0} + \ket{0'})/\sqrt{2}$. Then we carry out $N/2$ adiabatic processes to map population back and forth between $\ket{0}$ and $\ket{D}$, followed by a spin echo $\pi$ pulse on the clock transition, followed by $N/2$ adiabatic transfers. We then apply a final $\pi/2$  analysis pulse with varying phase and carry out a florescence measurement. As the phase is varied, we will see fringes in the measured population, just as in a standard Ramsey experiment. If there is any decoherence of the stored qubit, the amplitude of the fringes will decay. By fitting the population in $F = 1$ as a function of the phase of the final pulse, we can obtain the fidelity with which the qubit state is preserved. The decay of the fidelity with increasing $N$ is then measured in a similar way to before. This allows us to extract the average infidelity of the qubit mapping process, which is found to be $1 - \overline{\mathcal{F}} = 1.8(4) \times 10^{-4}$.

\section{Applications to other $d$-level systems}\label{sec:dlevels}
The technique described in this paper is general and can be applied to systems of arbitrary numbers of levels in a variety of quantum control applications. To illustrate this we provide two examples of potential applications in different quantum systems.

First we refer to the work of Liu et al. \cite{Liu}, who proposed a method to transfer the state of one $d$-level superconducting qudit to another in circuit QED. They illustrate their method in detail for the five-level case and show that it can be generalised to any number of levels. The method involves successively swapping over the population of different levels from one qudit to another via a cavity mode. By the end of step IV of their process (Figure 2 of \cite{Liu}) they have transferred the population of each individual state to the second qubit, but the states are in the wrong order. Therefore, in the final step of their process, Liu et al. apply a succession of pulses on different transitions within the qudit to rearrange the state populations so that they are in the exact reverse order compared to where they started. At this point the qudit transfer process is complete.

Here we show that, using our technique, a multi-level control method can instead be found to put the state populations back in their original order (not reversed) in a single step. Specifically, one must apply this four-level method to the top four levels of the second qudit (Figure 2 of \cite{Liu}) so as to reverse the order of their amplitudes. The required unitary matrix to carry out this operation is as follows:
\begin{equation}\label{eq:U}
U^{j=\frac{3}{2}} = \left(\begin{array}{cccc}
0 & 0 & 0 & 1 \\ 
0 & 0 & 1 & 0 \\ 
0 & 1 & 0 & 0 \\
1 & 0 & 0 & 0
\end{array} \right),
\end{equation}
and we are looking for a quantum control method to implement this unitary operation. This unitary transformation is (up to a global phase which can be easily accounted for by changing the phases of the other pulses in the sequence) equal to $e^{-i\pi J_x}$, which is a rotation of exactly the form we need to derive multi-level quantum control method using our technique. The equivalent two-level rotation is simply $e^{-i\pi S_x}$, which can be achieved by a variety of quantum control methods: for example a simple Rabi $\pi$-pulse or, if more robustness is required, more complex composite pulse or adiabatic schemes. The exact form of the control fields used to execute this transformation will depend on the exact control method used to implement the effective two level rotation. In general, for a single control field applied to a two-level system, the two-level Hamiltonian of equation \ref{eq:H1/2} must be transformed into a new four-level Hamiltonian using the spin-3/2 matrices. Physically, this Hamiltonian, which represents the desired quantum control method, will correspond to three different control fields on the four level-system, of varying Rabi frequencies and detunings.

Liu et al. discuss in their work how their method generalises to $d$ levels. Our four-level method also has a $d$-level equivalent which can reverse the populations of any number of states. One can verify this by noting that if you substitute $a = 0$, $b = i$ into equation \ref{eq:Ud} you obtain
\begin{equation}\label{eq:dNOT}
U^j_{rs} = i^{d+1}\delta_{d+1,r+s},
\end{equation}
where $d = 2j + 1$ is the number of levels and $\delta_{ij}$ is the Kronecker delta. This is indeed a unitary operation which reverses the order of the amplitudes for a $d$-level system.

Finally we consider the efficient Toffoli gate scheme discussed in Refs. \cite{Ralph, Lanyon}. Here the  three-level unitary operation
\begin{equation}\label{eq:Xa}
X_a = \left(\begin{array}{ccc}
0 & 0 & 1 \\ 
0 & 1 & 0 \\ 
1 & 0 & 0
\end{array} \right)
\end{equation}
is applied to a qutrit as part of the scheme. It is easy to verify that $X_a$ is in fact equal to $U^j_{rs}$ in equation \ref{eq:Xa} in the case where $d = 3$ (up to an irrelevant global phase), showing that this control operation is also amenable to the techniques described in this paper.

\bibliographystyle{naturemag}
\bibliography{ReportBib}

\begin{thebibliography}{10}
\expandafter\ifx\csname url\endcsname\relax
  \def\url#1{\texttt{#1}}\fi
\expandafter\ifx\csname urlprefix\endcsname\relax\def\urlprefix{URL }\fi
\providecommand{\bibinfo}[2]{#2}
\providecommand{\eprint}[2][]{\url{#2}}

\bibitem{Glaser}
\bibinfo{author}{Glaser, S.~J.} \emph{et~al.}
\newblock \bibinfo{title}{Training {S}chr{\"o}dinger’s cat: quantum optimal
  control}.
\newblock \emph{\bibinfo{journal}{Eur. Phys. J. D.}}
  \textbf{\bibinfo{volume}{69}}, \bibinfo{pages}{279} (\bibinfo{year}{2015}).
\newblock \urlprefix\url{https://arxiv.org/pdf/1508.00442.pdf}.

\bibitem{Mabuchi}
\bibinfo{author}{Mabuchi, H.} \& \bibinfo{author}{Navin, K.}
\newblock \bibinfo{title}{Principles and applications of control in quantum
  systems}.
\newblock \emph{\bibinfo{journal}{Int. J. Robust Nonlinear Control}}
  \textbf{\bibinfo{volume}{15}}, \bibinfo{pages}{647} (\bibinfo{year}{2005}).
\newblock
  \urlprefix\url{http://onlinelibrary.wiley.com/doi/10.1002/rnc.1016/pdf}.

\bibitem{Vandersypen}
\bibinfo{author}{Vandersypen, L. M.~K.} \& \bibinfo{author}{Chuang, I.~L.}
\newblock \bibinfo{title}{{NMR} techniques for quantum control and
  computation}.
\newblock \emph{\bibinfo{journal}{Rev. Mod. Phys.}}
  \textbf{\bibinfo{volume}{76}}, \bibinfo{pages}{1037} (\bibinfo{year}{2004}).
\newblock
  \urlprefix\url{http://journals.aps.org/rmp/pdf/10.1103/RevModPhys.76.1037}.

\bibitem{Fleischhauer}
\bibinfo{author}{Fleischhauer, M.}, \bibinfo{author}{Imamoglu, A.} \&
  \bibinfo{author}{Marangos, J.~P.}
\newblock \bibinfo{title}{Electromagnetically induced transparency: Optics in
  coherent media}.
\newblock \emph{\bibinfo{journal}{Rev. Mod. Phys.}}
  \textbf{\bibinfo{volume}{77}}, \bibinfo{pages}{633--673}
  (\bibinfo{year}{2005}).
\newblock \urlprefix\url{http://link.aps.org/doi/10.1103/RevModPhys.77.633}.

\bibitem{Kuhn}
\bibinfo{author}{Kuhn, A.}, \bibinfo{author}{Hennrich, M.} \&
  \bibinfo{author}{Rempe, G.}
\newblock \bibinfo{title}{Deterministic single-photon source for distributed
  quantum networking}.
\newblock \emph{\bibinfo{journal}{Phys. Rev. Lett.}}
  \textbf{\bibinfo{volume}{89}}, \bibinfo{pages}{067901}
  (\bibinfo{year}{2002}).
\newblock
  \urlprefix\url{http://link.aps.org/doi/10.1103/PhysRevLett.89.067901}.

\bibitem{Timoney}
\bibinfo{author}{Timoney, N.} \emph{et~al.}
\newblock \bibinfo{title}{Quantum gates and memory using microwave-dressed
  states}.
\newblock \emph{\bibinfo{journal}{Nature}} \textbf{\bibinfo{volume}{476}},
  \bibinfo{pages}{185--188} (\bibinfo{year}{2011}).
\newblock \urlprefix\url{http://dx.doi.org/10.1038/nature10319}.

\bibitem{Webster}
\bibinfo{author}{Webster, S.~C.}, \bibinfo{author}{Weidt, S.},
  \bibinfo{author}{Lake, K.}, \bibinfo{author}{McLoughlin, J.~J.} \&
  \bibinfo{author}{Hensinger, W.~K.}
\newblock \bibinfo{title}{Simple manipulation of a microwave dressed-state ion
  qubit}.
\newblock \emph{\bibinfo{journal}{Phys. Rev. Lett.}}
  \textbf{\bibinfo{volume}{111}}, \bibinfo{pages}{140501}
  (\bibinfo{year}{2013}).
\newblock
  \urlprefix\url{http://link.aps.org/doi/10.1103/PhysRevLett.111.140501}.

\bibitem{Kuklinski}
\bibinfo{author}{Kuklinski, J.~R.}, \bibinfo{author}{Gaubatz, U.},
  \bibinfo{author}{Hioe, F.~T.} \& \bibinfo{author}{Bergmann, K.}
\newblock \bibinfo{title}{Adiabatic population transfer in a three-level system
  driven by delayed laser pulses}.
\newblock \emph{\bibinfo{journal}{Phys. Rev. A}} \textbf{\bibinfo{volume}{40}},
  \bibinfo{pages}{6741--6744} (\bibinfo{year}{1989}).
\newblock \urlprefix\url{http://link.aps.org/doi/10.1103/PhysRevA.40.6741}.

\bibitem{Rangelov}
\bibinfo{author}{Rangelov, A.~A.} \emph{et~al.}
\newblock \bibinfo{title}{Stark-shift-chirped rapid-adiabatic-passage technique
  among three states}.
\newblock \emph{\bibinfo{journal}{Phys. Rev. A}} \textbf{\bibinfo{volume}{72}},
  \bibinfo{pages}{053403} (\bibinfo{year}{2005}).
\newblock \urlprefix\url{http://link.aps.org/doi/10.1103/PhysRevA.72.053403}.

\bibitem{Broers}
\bibinfo{author}{Broers, B.}, \bibinfo{author}{van Linden van~den Heuvell,
  H.~B.} \& \bibinfo{author}{Noordam, L.~D.}
\newblock \bibinfo{title}{Efficient population transfer in a three-level ladder
  system by frequency-swept ultrashort laser pulses}.
\newblock \emph{\bibinfo{journal}{Phys. Rev. Lett.}}
  \textbf{\bibinfo{volume}{69}}, \bibinfo{pages}{2062--2065}
  (\bibinfo{year}{1992}).
\newblock \urlprefix\url{http://link.aps.org/doi/10.1103/PhysRevLett.69.2062}.

\bibitem{Melinger}
\bibinfo{author}{Melinger, J.~S.}, \bibinfo{author}{Gandhi, S.~R.},
  \bibinfo{author}{Hariharan, A.}, \bibinfo{author}{Tull, J.~X.} \&
  \bibinfo{author}{Warren, W.~S.}
\newblock \bibinfo{title}{Generation of narrowband inversion with broadband
  laser pulses}.
\newblock \emph{\bibinfo{journal}{Phys. Rev. Lett.}}
  \textbf{\bibinfo{volume}{68}}, \bibinfo{pages}{2000--2003}
  (\bibinfo{year}{1992}).
\newblock \urlprefix\url{http://link.aps.org/doi/10.1103/PhysRevLett.68.2000}.

\bibitem{Vitanov}
\bibinfo{author}{Vitanov, N.~V.}, \bibinfo{author}{Halfmann, T.},
  \bibinfo{author}{Shore, B.~W.} \& \bibinfo{author}{Bergmann, K.}
\newblock \bibinfo{title}{Laser-induced population transfer by adiabatic
  passage techniques}.
\newblock \emph{\bibinfo{journal}{Annual Review of Physical Chemistry}}
  \textbf{\bibinfo{volume}{52}}, \bibinfo{pages}{763--809}
  (\bibinfo{year}{2001}).

\bibitem{Khaneja}
\bibinfo{author}{Khaneja, N.}, \bibinfo{author}{Reiss, T.},
  \bibinfo{author}{Kehlet, C.}, \bibinfo{author}{Schulte-Herbr{\"u}ggen, T.} \&
  \bibinfo{author}{Glaser, S.~J.}
\newblock \bibinfo{title}{Optimal control of coupled spin dynamics: design of
  {NMR} pulse sequences by gradient ascent algorithms}.
\newblock \emph{\bibinfo{journal}{Journal of Magnetic Resonance}}
  \textbf{\bibinfo{volume}{172}}, \bibinfo{pages}{296 -- 305}
  (\bibinfo{year}{2005}).
\newblock
  \urlprefix\url{http://www.sciencedirect.com/science/article/pii/S1090780704003696}.

\bibitem{Majorana}
\bibinfo{author}{Majorana, E.}
\newblock \bibinfo{title}{Atomi orientati in campo magnetico variabile}.
\newblock \emph{\bibinfo{journal}{Il Nuovo Cimento (1924-1942)}}
  \textbf{\bibinfo{volume}{9}}, \bibinfo{pages}{43--50} (\bibinfo{year}{1932}).
\newblock \urlprefix\url{http://dx.doi.org/10.1007/BF02960953}.

\bibitem{Bloch}
\bibinfo{author}{Bloch, F.} \& \bibinfo{author}{Rabi, I.~I.}
\newblock \bibinfo{title}{Atoms in variable magnetic fields}.
\newblock \emph{\bibinfo{journal}{Rev. Mod. Phys.}}
  \textbf{\bibinfo{volume}{17}}, \bibinfo{pages}{237--244}
  (\bibinfo{year}{1945}).
\newblock \urlprefix\url{http://link.aps.org/doi/10.1103/RevModPhys.17.237}.

\bibitem{Hioe}
\bibinfo{author}{Hioe, F.~T.}
\newblock \bibinfo{title}{N-level quantum systems with {SU(2)} dynamic
  symetry}.
\newblock \emph{\bibinfo{journal}{J. Opt. Soc. Am. B}}
  \textbf{\bibinfo{volume}{4}}, \bibinfo{pages}{1327--1332}
  (\bibinfo{year}{1987}).
\newblock
  \urlprefix\url{https://www.osapublishing.org/josab/abstract.cfm?uri=josab-4-8-13273}.

\bibitem{CookShore}
\bibinfo{author}{Cook, R.~J.} \& \bibinfo{author}{Shore, B.~W.}
\newblock \bibinfo{title}{Coherent dynamics of n-level atoms and molecules.
  {III}. an analytically soluble periodic case}.
\newblock \emph{\bibinfo{journal}{Phys. Rev. A}} \textbf{\bibinfo{volume}{20}},
  \bibinfo{pages}{539--544} (\bibinfo{year}{1979}).
\newblock
  \urlprefix\url{https://journals.aps.org/pra/abstract/10.1103/PhysRevA.20.539}.

\bibitem{Genov}
\bibinfo{author}{Genov, G.~T.}, \bibinfo{author}{Torosov, B.~T.} \&
  \bibinfo{author}{Vitanov, N.~V.}
\newblock \bibinfo{title}{Optimized control of multistate quantum systems by
  composite pulse sequences}.
\newblock \emph{\bibinfo{journal}{Phys. Rev. A}} \textbf{\bibinfo{volume}{84}},
  \bibinfo{pages}{063413} (\bibinfo{year}{2011}).
\newblock
  \urlprefix\url{https://journals.aps.org/pra/pdf/10.1103/PhysRevA.84.063413}.

\bibitem{Bergeman}
\bibinfo{author}{Bergeman, T.~H.}, \bibinfo{author}{McNicholl, P.},
  \bibinfo{author}{Kycia, J.}, \bibinfo{author}{Metcalf, H.} \&
  \bibinfo{author}{Balazs, N.~L.}
\newblock \bibinfo{title}{Quantized motion of atoms in a quadrupole
  magnetostatic trap}.
\newblock \emph{\bibinfo{journal}{J. Opt. Soc. Am. B}}
  \textbf{\bibinfo{volume}{6}}, \bibinfo{pages}{2249--2256}
  (\bibinfo{year}{1989}).
\newblock
  \urlprefix\url{http://josab.osa.org/abstract.cfm?URI=josab-6-11-2249}.

\bibitem{Bauch}
\bibinfo{author}{Bauch, A.} \& \bibinfo{author}{Schr{\"o}der, R.}
\newblock \bibinfo{title}{Frequency shifts in a cesium atomic clock due to
  majorana transitions}.
\newblock \emph{\bibinfo{journal}{Ann. Phys.}} \textbf{\bibinfo{volume}{505}},
  \bibinfo{pages}{421} (\bibinfo{year}{1993}).
\newblock
  \urlprefix\url{http://onlinelibrary.wiley.com/doi/10.1002/andp.19935050502/pdf}.

\bibitem{McLoughlin2}
\bibinfo{author}{McLoughlin, J.~J.} \emph{et~al.}
\newblock \bibinfo{title}{Versatile ytterbium ion trap experiment for operation
  of scalable ion-trap chips with motional heating and transition-frequency
  measurements}.
\newblock \emph{\bibinfo{journal}{Phys. Rev. A}} \textbf{\bibinfo{volume}{83}},
  \bibinfo{pages}{013406} (\bibinfo{year}{2011}).
\newblock \urlprefix\url{http://link.aps.org/doi/10.1103/PhysRevA.83.013406}.

\bibitem{Lake}
\bibinfo{author}{Lake, K.} \emph{et~al.}
\newblock \bibinfo{title}{Generation of spin-motion entanglement in a trapped
  ion using long-wavelength radiation}.
\newblock \emph{\bibinfo{journal}{Phys. Rev. A}} \textbf{\bibinfo{volume}{91}},
  \bibinfo{pages}{012319} (\bibinfo{year}{2015}).
\newblock \urlprefix\url{http://link.aps.org/doi/10.1103/PhysRevA.91.012319}.

\bibitem{Randall}
\bibinfo{author}{Randall, J.} \emph{et~al.}
\newblock \bibinfo{title}{Efficient preparation and detection of microwave
  dressed-state qubits and qutrits with trapped ions}.
\newblock \emph{\bibinfo{journal}{Phys. Rev. A}} \textbf{\bibinfo{volume}{91}},
  \bibinfo{pages}{012322} (\bibinfo{year}{2015}).
\newblock \urlprefix\url{https://link.aps.org/doi/10.1103/PhysRevA.91.012322}.

\bibitem{Weidt2}
\bibinfo{author}{Weidt, S.} \emph{et~al.}
\newblock \bibinfo{title}{Ground-state cooling of a trapped ion using
  long-wavelength radiation}.
\newblock \emph{\bibinfo{journal}{Phys. Rev. Lett.}}
  \textbf{\bibinfo{volume}{115}}, \bibinfo{pages}{013002}
  (\bibinfo{year}{2015}).
\newblock
  \urlprefix\url{http://link.aps.org/doi/10.1103/PhysRevLett.115.013002}.

\bibitem{Weidt}
\bibinfo{author}{Weidt, S.} \emph{et~al.}
\newblock \bibinfo{title}{Trapped-ion quantum logic with global radiation
  fields}.
\newblock \emph{\bibinfo{journal}{Phys. Rev. Lett.}}
  \textbf{\bibinfo{volume}{117}}, \bibinfo{pages}{220501}
  (\bibinfo{year}{2016}).
\newblock \urlprefix\url{http://www.sussex.ac.uk/physics/iqt/globalfields.pdf}.

\bibitem{Bermudez3}
\bibinfo{author}{Bermudez, A.}, \bibinfo{author}{Jelezko, F.},
  \bibinfo{author}{Plenio, M.~B.} \& \bibinfo{author}{Retzker, A.}
\newblock \bibinfo{title}{Electron-mediated nuclear-spin interactions between
  distant nitrogen-vacancy centers}.
\newblock \emph{\bibinfo{journal}{Phys. Rev. Lett}}
  \textbf{\bibinfo{volume}{107}}, \bibinfo{pages}{150503}
  (\bibinfo{year}{2011}).
\newblock
  \urlprefix\url{http://journals.aps.org/prl/abstract/10.1103/PhysRevLett.107.150503}.

\bibitem{Baumgart}
\bibinfo{author}{Baumgart, I.}, \bibinfo{author}{Cai, J.-M.},
  \bibinfo{author}{Retzker, A.}, \bibinfo{author}{Plenio, M.~B.} \&
  \bibinfo{author}{Wunderlich, C.}
\newblock \bibinfo{title}{Ultrasensitive magnetometer using a single atom}.
\newblock \emph{\bibinfo{journal}{Phys. Rev. Lett.}}
  \textbf{\bibinfo{volume}{116}}, \bibinfo{pages}{240801}
  (\bibinfo{year}{2016}).
\newblock
  \urlprefix\url{http://link.aps.org/doi/10.1103/PhysRevLett.116.240801}.

\bibitem{Landau}
\bibinfo{author}{Landau, L.}
\newblock \bibinfo{title}{On the theory of transfer of energy at collisions
  {II}}.
\newblock \emph{\bibinfo{journal}{Phys. Z. Sowjetunion}}
  \textbf{\bibinfo{volume}{2}}, \bibinfo{pages}{46} (\bibinfo{year}{1932}).

\bibitem{Zener}
\bibinfo{author}{Zener, C.}
\newblock \bibinfo{title}{Non-adiabatic crossing of energy levels}.
\newblock \emph{\bibinfo{journal}{Proc. R. Soc. London A}}
  \textbf{\bibinfo{volume}{137}}, \bibinfo{pages}{696} (\bibinfo{year}{1932}).

\bibitem{Blackman}
\bibinfo{author}{Blackman, R.~B.} \& \bibinfo{author}{Tukey, J.~W.}
\newblock \bibinfo{title}{Particular pairs of windows}.
\newblock In \emph{\bibinfo{booktitle}{The Measurement of Power Spectra, From
  the Point of View of Communications Egineering}}, \bibinfo{pages}{98--99}
  (\bibinfo{publisher}{New York: Dover}, \bibinfo{year}{1959}).

\bibitem{Wimperis}
\bibinfo{author}{Wimperis, S.}
\newblock \bibinfo{title}{Broadband, narrowband, and passband composite pulses
  for use in advanced {NMR} experiments}.
\newblock \emph{\bibinfo{journal}{Journal of Magnetic Resonance, Series A}}
  \textbf{\bibinfo{volume}{109}}, \bibinfo{pages}{221 -- 231}
  (\bibinfo{year}{1994}).
\newblock
  \urlprefix\url{http://www.sciencedirect.com/science/article/pii/S1064185884711594}.

\bibitem{Note1}
\bibinfo{note}{A recent study has suggested that the true infidelities may
  actually be somewhat lower than this, as infidelities due to coherent effects
  may be overestimated by this method \cite {Vitanov3}. However, as we believe
  our infidelities to be dominated by incoherent effects such as decoherence,
  this effect may not be significant.}

\bibitem{Lekitsch}
\bibinfo{author}{Lekitsch, B.} \emph{et~al.}
\newblock \bibinfo{title}{Blueprint for a microwave trapped ion quantum
  computer}.
\newblock \emph{\bibinfo{journal}{Sci. Adv.}} \textbf{\bibinfo{volume}{3}},
  \bibinfo{pages}{e1601540} (\bibinfo{year}{2017}).
\newblock \urlprefix\url{http://www.sussex.ac.uk/physics/iqt/blueprint.pdf}.

\bibitem{Torosov}
\bibinfo{author}{Torosov, B.~T.} \& \bibinfo{author}{Vitanov, N.~V.}
\newblock \bibinfo{title}{Evolution of superpositions of quantum states through
  a level crossing}.
\newblock \emph{\bibinfo{journal}{Phys. Rev. A}} \textbf{\bibinfo{volume}{84}},
  \bibinfo{pages}{063411} (\bibinfo{year}{2011}).
\newblock \urlprefix\url{http://link.aps.org/doi/10.1103/PhysRevA.84.063411}.

\bibitem{Liu}
\bibinfo{author}{Liu, T.} \emph{et~al.}
\newblock \bibinfo{title}{Transferring arbitrary d-dimensional quantum states
  of a superconducting transmon qudit in circuit qed}.
\newblock \emph{\bibinfo{journal}{Scientific Reports}}
  \textbf{\bibinfo{volume}{7}}, \bibinfo{pages}{7039} (\bibinfo{year}{2017}).
\newblock
  \urlprefix\url{https://www.nature.com/articles/s41598-017-07225-5.pdf}.

\bibitem{Ralph}
\bibinfo{author}{Ralph, T.~C.}, \bibinfo{author}{Resch, K.~J.} \&
  \bibinfo{author}{Gilchrist, A.}
\newblock \bibinfo{title}{Efficient toffoli gates using qudits}.
\newblock \emph{\bibinfo{journal}{Phys. Rev. A}} \textbf{\bibinfo{volume}{75}},
  \bibinfo{pages}{022313} (\bibinfo{year}{2007}).
\newblock
  \urlprefix\url{https://journals.aps.org/pra/pdf/10.1103/PhysRevA.75.022313}.

\bibitem{Lanyon}
\bibinfo{author}{Lanyon, B.~P.} \emph{et~al.}
\newblock \bibinfo{title}{Simplifying quantum logic using higher-dimensional
  hilbert spaces}.
\newblock \emph{\bibinfo{journal}{Nature Phys.}} \textbf{\bibinfo{volume}{5}},
  \bibinfo{pages}{134--140} (\bibinfo{year}{2008}).
\newblock \urlprefix\url{https://www.nature.com/articles/nphys1150.pdf}.

\bibitem{Vitanov3}
\bibinfo{author}{Vitanov, N.~V.}
\newblock \bibinfo{title}{Relations between the single-pass and double-pass
  transition probabilities in quantum systems with two and three states}.
\newblock \emph{\bibinfo{journal}{Phys. Rev. A}} \textbf{\bibinfo{volume}{97}},
  \bibinfo{pages}{053409} (\bibinfo{year}{2018}).
\newblock
  \urlprefix\url{https://journals.aps.org/pra/pdf/10.1103/PhysRevA.97.053409}.

\end{thebibliography}

\end{document}